\documentclass[journal,10pt]{IEEEtran}
\IEEEoverridecommandlockouts

\usepackage[utf8]{inputenc}
\usepackage[T1]{fontenc} 
\usepackage[american]{babel}
\usepackage[flushleft]{caption}
\usepackage{graphicx}
\usepackage{subfigure}
\usepackage{listings}
\usepackage{float}
\usepackage{stfloats}
\usepackage{amsmath,amssymb,exscale}
\usepackage{blindtext, graphicx}
\usepackage{verbatim}
\usepackage{algorithm}
\usepackage{algorithmicx}
\usepackage{algpseudocode}
\usepackage{fancyvrb}
\usepackage{bera}
\usepackage{amsmath}
\usepackage{verbatim}
\usepackage{mathtools}
\usepackage{lipsum}
\usepackage{bm}
\usepackage{geometry}
\usepackage{color}
\usepackage{float}
\usepackage{lettrine}
\usepackage{pifont} 

\usepackage{times}
\usepackage{cite}
\usepackage{scalerel}
\usepackage{tikz}
\usepackage[bookmarks=false]{hyperref}

\usetikzlibrary{svg.path}
\usepackage{authblk}
\title{ When UAVs Meet ISAC: Real-Time Trajectory Design for Secure Communications}
  \author{
    \IEEEauthorblockN{Jun Wu, \IEEEmembership{Student Member, IEEE,}  Weijie Yuan, \IEEEmembership{Member, IEEE}, and Lajos Hanzo, \IEEEmembership{Life Fellow, IEEE}
  }
  \thanks{This work was supported in part by the National Natural Science Foundation of China under Grants 62101232 and 62201242; in part by Guangdong Provincial Natural Science Foundation under Grant 2022A1515011257; in part by Shenzhen Science and Technology Program under Grant JCYJ20220530114412029.  L. Hanzo would like to acknowledge the Engineering and Physical Sciences Research Council projects EP/W016605/1 and EP/P003990/1 (COALESCE) as well as of the European Research Council’s Advanced Fellow Grant QuantCom (Grant No. 789028).  (Corresponding author: Weijie Yuan)

  J. Wu and W. Yuan are with the Department of Electrical and Electronic Engineering, Southern University of Science and
  Technology, Shenzhen 518055, China (email: wuj2021@mail.sustech.edu.cn; yuanwj@sustech.edu.cn).
  
 L. Hanzo is with the School of Electronics and Computer Science, University of Southampton, Southampton SO17 1BJ, UK. (e-mail: lh@ecs.soton.ac.uk).
  }
  }
\begin{document}
\maketitle
\begin{abstract}
The real-time unmanned aerial vehicle (UAV)  trajectory design of secure integrated sensing and communication (ISAC) is optimized. In particular, the UAV serves both as a downlink transmitter and a radar receiver. The legitimate user (Bob) roams on ground through a series of unknown locations, while the eavesdropper moves following a fixed known trajectory. To maximize the real-time secrecy rate, we propose an extended Kalman filtering (EKF)-based method for tracking and predicting Bob's location at the UAV based on the delay measurements extracted from the sensing echoes. We then formulate a non-convex real-time trajectory design problem and develop an efficient iterative algorithm for finding a near optimal solution. Our numerical results demonstrate that the proposed algorithm is capable of accurately tracking Bob and strikes a compelling legitimate vs. leakage rate trade-off.
\end{abstract}
\vspace{-0.2cm}
\section{Introduction }
\lettrine[lines=2]{G}{iven} by the high flexibility and predominantly line-of-sight (LoS) mature of air-to-ground propagation links, unmanned aerial vehicles (UAVs) are capable of providing reliable communication services in rural, disaster, and hot-spot areas in next generation networks\cite{networks}. In contrast to traditional terrestrial communication systems, UAVs are capable of dynamically adjusting their coverage areas to serve as mobile relays or aerial base stations (BSs) \cite{func}. To meet the demand for high rate transmission in next generation systems, the authors of \cite{deployment} investigated the deployment of UAV relays in the presence of malfunctioning base stations and maximized the capacity of the relay network. However, stationary relays do not have the high flexibility of UAVs. Furthermore, UAV communications are highly susceptible to potential eavesdropping due to their LoS-dominated communication channels \cite{secure}. Hence, there is an emerging trend to design secure UAV-based communication schemes. For instance, the authors of \cite{singleuser} investigated the maximum secrecy rate of a single user relying on joint trajectory design and power allocation. Furthermore, the energy efficiency maximization of secure communication systems was considered in \cite{yuanxincai}, while supporting multiple users. However, all these contributions assumed that both the ground users as well as eavesdroppers are static on the ground, while, in practice the ground users are motive, hence imposing challenges on UAV-aided secure communication systems.

To enhance the communication performance in practical applications in the face of user mobility, the UAVs are required to keep track of the real-time locations of ground users \cite{meng2022uav}. Traditionally, UAV-based localization and tracking schemes tend to rely on the Global Navigation Satellite System (GNSS) and/or video sensors. For example, equipped with the measurements of UAV location and camera  angles, the authors of \cite{vision} proposed a vision-based localization method by exploiting the pixel-based location estimate of the target in an image. However, such vision sensor-based methods are likely to suffer localization performance degradation due to the environmental variations. Moreover, attaching vision sensors to UAVs will increase their sizes and result in additional power consumption, which is undesirable for UAVs having limited onboard battery capacity. 
\begin{figure}[t]
  \centering
  \includegraphics[scale=0.8]{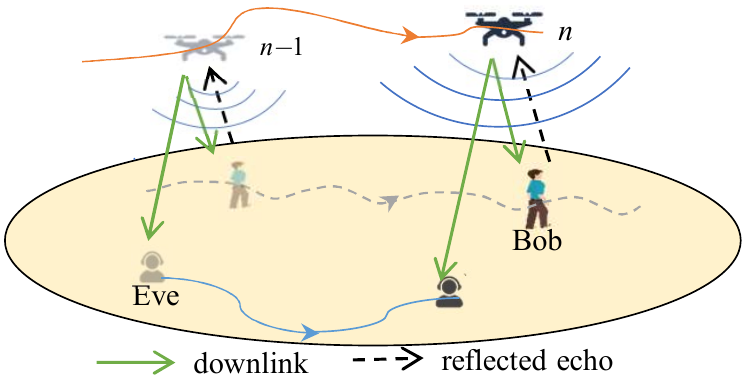}
  \captionsetup{font={scriptsize},singlelinecheck=off}
  \caption{A UAV-enabled ISAC system. }\label{fig1}
  \vspace{-0.5cm}
\end{figure}
To simultaneously support UAV-based communication and positioning services at a low overhead, there is a novel technology integrating both sensing and communication functionalities using unified hardware and signal waveforms \cite{lu2022degrees,du}. This so-called integrated sensing and communication (ISAC) technology \cite{ISAC1} allows traditionally independent sensing and communication systems to seamlessly share both their wireless infrastructure and spectral resources for significantly enhancing their spectrum-, energy-, and hardware-efficiency. Recently, some research efforts have been devoted to ISAC systems relying on UAV platforms. For example, Meng \emph{et.al} \cite{mengwcl2022uav} proposed a novel UAV-enabled integrated periodic sensing and communication mechanism to strike a trade-off between sensing and communication. As a step forward, the authors of \cite{mengkaitao2022throughput} maximized the achievable communication rate, while meeting the sensing frequency and beam pattern gain requirement.

 In this context, we solve the problem of simultaneous ground user tracking and secure communications for a UAV-based system. In particular, a legitimate user (Bob) moves through unknown locations and an eavesdropper (Eve) follows a fixed trajectory\footnote{These assumptions are typically satisfied in applications, where Eve moves along a preset path for eavesdropping. The more generalized scenario, where Eve has an unknown trajectory which has to be predicted will be considered in our future work.}. 
 For accurately tracking Bob, we first propose an extended Kalman filtering (EKF) framework, which relies on the delay measurements extracted from the ISAC echoes. Furthermore, based on the predicted location, we formulate a weighted non-convex trajectory design problem for supporting flexible secure communications performance for meeting the diverse requirements of various UAV-based applications. We then propose an efficient and rapidly converging iterative algorithm for solving the resultant non-convex optimization problem via the popular successive convex approximation (SCA) technique. Our simulation results show that the proposed algorithm efficiently tracks Bob and maximizes the real-time secrecy rate in the presence of Eve.
 
 The rest of this paper is organized as follows. Section \uppercase\expandafter{\romannumeral2} introduces the UAV-based system model. Section \uppercase\expandafter{\romannumeral3} formulates the real-time trajectory design problem, while the proposed solutions are derived in Section \uppercase\expandafter{\romannumeral4}. Our simulation results are provided in Section \uppercase\expandafter{\romannumeral5}, while Section \uppercase\expandafter{\romannumeral6} concludes this paper.

 \textit{Notations:} The $M$-dimensional vector space is denoted as $\mathbb{R}^{M\times 1}$. We use $\Vert\cdot\Vert$ and $[\cdot]^ {\rm T}$ to denote the vector norm and the transposition operation, respectively. We use $\mathcal{N} (\mu,v)$ to denote a Gaussian distribution of mean $\mu$ and variance $v$. For a time-dependent function $x(t)$, the first-order derivatives with respect to time $t$ are denoted as $\dot x(t)$. We use $\mathbf{I}$ and $\hat{x}$ to denote the identity matrix and the estimated value of $x$, respectively. We use diag($\cdot$) to represent a diagonal matrix.

\section{System Model}
As shown in Fig. \ref*{fig1}, we consider a UAV-aided ISAC system where the UAV acts as a downlink communication transmitter and a radar receiver to serve Bob in the presence of Eve. Assume that the UAV hovers in the sky with a total flight period of $T$, which can be divided into $N$ time slots (TSs), contained in the set $\mathcal{F}=\left\{1,2,..., N\right\}$. The duration between two consecutive TSs is denoted as $\Delta t>0$. We consider a three-dimensional Cartesian coordinate system. At the $n$-th TS, the time-varying horizontal coordinate of the UAV is denoted as $\mathbf{q}[n]=[{x_q}[n],{y_q}[n]]^{\text{T}} \in \mathbb{R}^{2\times 1}$ with a constant altitude $H$. Since Bob and Eve are on the ground at zero altitudes, the coordinates of Bob and Eve at the $n$-th TS can be expressed as $\mathbf{b}[n]= [x_b[n],y_b [n]]^\text{T} \in \mathbb{R}^{2\times 1}$, $\mathbf{w}[n] = [x_w[n],y_w[n]]^\text{T}\in \mathbb{R}^{2\times 1}$, respectively.
\subsection{Sensing Model}
 The UAV aims for tracking Bob for providing improved communication performance. In each TS, the UAV transmits its information-bearing signal $c(t)$ to Bob. Since the UAV's sensing capability, the echoes reflected by Bob and received at the UAV can be written as
\begin{align}
  r(t)=A \beta c\left(t-\tau\right) e^{j 2 \pi f_{d} t}+z_r(t),
\end{align}
where $A$, $\beta$, $\tau$, and $f_d$ represent the echoes amplitude, the reflection coefficient, the round-trip signaling delay, and the Doppler frequency, respectively. While $z_r(t)$ represents the additive white Gaussian noise (AWGN) process. Let us express the distance from the UAV to Bob and the UAV to Eve at the $n$-th TS as
\begin{align}
  d_b[n]&=\sqrt{H^2+\Vert\mathbf{q}[n]-\mathbf{b}[n]\Vert^2}
 \quad  \text{and} \\ d_w[n]&=\sqrt{H^2+\Vert\mathbf{q}[n]-\mathbf{w}[n]\Vert^2},
\end{align} 
respectively\footnote{Note that the identification of Bob is essential for the success of the proposed real-time sensing-aided secure communications design. 
Although how to identify Bob is not in the main scope of this manuscript, our proposed framework can efficiently distinguish Bob by solving the corresponding data association problem \cite{4797815} using a typical radar cross section (RCS) size, and the specific trajectories of the objects. We can also leverage the periodic location feedback mechanism discussed in Sec.V to ascertain the identity of Bob.}.  In practice, it is essential to acknowledge the potential existence of unwanted echoes that are reflected by other terrestrial objects. Effective UAV-based filtering and deconvolutional methods have been extensively explored and employed for suppressing clutter interference \cite{jing,hu2019}. Given the signal propagation delay $\tau[n]$  at the $n$-th TS, the range between the UAV and Bob is given by $\frac{ c\tau[n]}{2}$, where $c$ represents the signal propagation speed. With the assumption of Gaussian measurement noise, the measured range $\hat{d}_b[n]$ is written as
\begin{align}
\hat{d}_b[n]&=d_b[n]+z_{d}, \label{dis} \
\end{align}
where $z_{d}$ is the corresponding noise term obeying the Gaussian distribution of $\mathcal{N}(0,\sigma_{d}^2)$. 

\subsection{Communication Model}
As for communications, since we consider a free-space propagation scenario, the channel gains corresponding to the UAV-Bob link and the UAV-Eve link follow the classic inverse second power path-loss model \cite{yuanxincai}, which can be expressed as  
   \begin{align}
     h_b[n]&=\frac{\rho_0 }{{H}^2+\Vert\mathbf{q}[n]-\mathbf{b}[n]\Vert^2},\\ 
   \text{and} \quad h_w[n]&=\frac{\rho_0 }{H^2+\Vert\mathbf{q}[n]-\mathbf{w}[n]\Vert^2},  
    \end{align}
  where $\rho_0$ denotes the reference channel's power gain at a unit distance. Hence, we assume the transmitted power to be approximately a constant value of $p_0$. This is mainly because the transmit power is far lower than the power dissipated by the propulsion of flying and hovering \cite{jing} and its variation is negligible. Consequently, the achievable communication rate of Bob at the $n$-th TS is given by
  \begin{align}
    R_{b}[n] =B\log_2(1+\frac{p_0h_b[n]}{\sigma^2}),    \label{RB}
  \end{align}
where $B$ and $\sigma^2$ are the channel's bandwidth and the noise power at Bob, respectively. Similarly, the leakage data rate at Eve at the $n$-th TS is given by
\begin{align}
 R_{w}[n] =B\log_2(1+\frac{p_0h_w[n]}{\sigma_e^2}),
  \label{RE}
\end{align}  
where $\sigma_e^2$ represents the noise power at Eve. According to (\ref*{RB}) and (\ref*{RE}), the achievable secrecy rate at the $n$-th TS becomes 
 \begin{align}
  R_s[n]= R_b[n]-R_w[n].
  \label{srate}
 \end{align}

\subsection{State Evolution Model}
We aim for tracking Bob's motion state at each TS, which is determined by Bob's kinematic equations. Due to the constraints of the road shape and the environment, for convenience, we assume that Bob is moving at an approximately constant speed in any pair of consecutive TSs. At the $n$-th TS, Bob's velocity is denoted as $\mathbf{v}[n]=[v_x[n],v_y[n]]^\text{T}$, where $v_x[n]$ and $v_y[n]$ are the speed along the $x$-axis and the $y$-axis in Cartesian coordinates, respectively. Hence, relying on the parameters at the ($n-1$)-st TS, the state evolution model for Bob's location  $[x_b[n],y_b[n]]^\text{T}$ and velocity $[v_x[n],v_y[n]]^\text{T}$ can be summarized as follows:
\begin{align}
   \label{locax} x_b[n]&=x_b[n-1]+v_{x}[n-1] \Delta t+\omega_x, \\  
    v_{x}[n]&=v_{x}[n-1]+\omega_{v_x}, \\ 
    y_b[n]&=y_b[n-1]+v_{y}[n-1] \Delta t+\omega_y, \\ 
    v_{y}[n]&=v_{y}[n-1]+\omega_{v_y} \label{locay},
\end{align}
 where $\omega_x$, $\omega_{v_x}$, $\omega_y$, and $\omega_{v_y}$ represent the corresponding transition noise terms having the distributions of $\mathcal{N}(0,\sigma_x^2)$, $\mathcal{N}(0,\sigma_{v_x}^2)$,  $\mathcal{N}(0,\sigma_y^2)$, and $\mathcal{N}(0,\sigma_{v_y}^2)$, respectively.
\vspace{-0.2cm}
\section{Problem Formulation}
 \par  Our goal is to maximize the real-time secrecy rate as a function of the UAV trajectory. Since the UAV's maximum speed $V_{max}$ is finite, the maximum travel distance is constrained as 
   \begin{align}
     \|\mathbf{q}[n]-\mathbf{q}[n-1]\Vert \leq V_{max} \Delta t, \quad \forall n\in \mathcal{F}. \label{velocity}
    \end{align}
  In practice,  the UAV will only fly in a rectangular area $\mathcal{D}$ having dimensions of $L_x$ and $L_y$. Thus, the UAV trajectory has to satisfy 
    \begin{align} \label{area}
      0\le x_q[n] \le L_x,\quad \forall n\in \mathcal{F}.\\
      0\le y_q[n] \le L_y,\quad \forall n\in \mathcal{F}.\label{area1}
    \end{align}
 Our optimization problem then can be formulated as
  \begin{subequations}
    \label{optt}
   \begin{align}
\max_{\mathbf{q}[n]} 
\nonumber \quad & \alpha B\log_2\left(1+\frac{p_0\rho_0}{\sigma^2\left(H^2+\Vert\mathbf{q}[n]-\mathbf{{b}}[n]\Vert^2\right)}\right) \\    \label{optt1}
& -(1-\alpha)B\log_2\left(1+\frac{p_0\rho_0}{\sigma_e^2\left(H^2+\Vert\mathbf{q}[n]-\mathbf{{w}}[n]\Vert^2\right)}\right)\\
\text{s.t.}\ \quad &  (\ref*{velocity}),(\ref*{area}),(\ref*{area1}),
   \end{align}
  \end{subequations}
 where $\alpha$ is a weighting factor taking values between 0 and 1 to achieve a flexible secure communication performance. A higher value of  $\alpha$ means that increasing the communications rate for Bob is more important than avoding information leakage. According to the above equations, it is seen that maximizing the secrecy rate requires the real-time design of the UAV's trajectory, which in turn requires the knowledge of Bob's location.
 \vspace*{-0.3cm}
 
\section{ Proposed Algorithm}
In Section \uppercase\expandafter{\romannumeral3}, we formulated the real-time UAV trajectory design as problem (\ref*{optt}), which is non-convex and hence it cannot be solved by conventional convex optimization methods. In this section, we first present an EKF-based algorithm conceived for tracking and predicting Bob's real-time location. Then, an iterative algorithm is proposed for solving problem (\ref*{optt}).
\subsection{EFK-Based Bob Tracking}
  \par A core contribution of our work is to treat the trajectory design as an on-line optimization, which is totally different from the global path planning in \cite{secure} and requires us to track Bob's motion state including location and velocity at each TS. The sophisticated Kalman filtering (KF) is a popular technique, which is often adopted for solving  estimation problems. But since the range measurements in (\ref*{dis}) are nonlinear, the classic KF can not be directly adopted. To circumvent this problem, we resort to the EKF algorithm, which has been widely used for solving nonlinear estimation problems.
 Let us define the state variables as $\mathbf{x}[n]=[x_b[n],\dot x_b[n],y_b[n],\dot y_b[n]]^\text{T}$. Accordingly, the discrete time dynamic models can be written in compact forms as
 \begin{align}
 \label{model}
  \left\{\begin{array}{l}
    \text {Evolution Model: } \mathbf{x}[n]=\boldsymbol{\Phi}[n | n-1]\mathbf{x}[n-1]+ \bm{\omega}[n-1], \\
    \text {Measurement Model: } \hat{d}_b[n]=h\left(\mathbf{x}[n]\right)+z_{d}[n],
    \end{array}\right.
 \end{align}
  where $\boldsymbol{\Phi}[n | n-1]$ is the linear state evolution matrix given by 
  \begin{align}
  \boldsymbol{\Phi}[n|n-1]=\left[\begin{array}{cccc}
      1 & \Delta t & 0 & 0 \\
      0 & 1 & 0 & 0 \\
      0 & 0 & 1 & \Delta t \\
      0 & 0 & 0 & 1
      \end{array}\right].
  \end{align}
 The vector $\bm{\omega}=[ \omega_x, \omega_{v_x}, \omega_y, \omega_{v_y}]^\mathrm{T}$ is the transition noise vector having the covariance matrix of
 \begin{align}
  \mathbf{Q}_{\omega}=\text{diag}( \sigma_x^2, \sigma_{v_x}^2, \sigma_y^2, \sigma_{v_y}^2).
 \end{align}
 In (\ref*{model}), $h(\cdot)$ is the nonlinear observation function defined in (\ref*{dis}). We then harness the Jacobian matrix of $h(\mathbf{x})$ to linearize the measurements shown as
\begin{align}
  \frac{\partial h}{\partial \mathbf{x}}&= \left[\begin{array}{cccc}
    \frac{\partial h}{\partial x} & \frac{\partial h}{\partial v_x} & \frac{\partial h}{\partial y}& \frac{\partial h}{\partial v_y} 
    \end{array}\right] \nonumber\\& =  \left[\begin{array}{cccc}
    \frac{x_b[n]-x_q[n]}{d_b[n]} &0 &\frac{y_b[n]-y_q[n]}{d_b[n]} &0  
    \end{array}\right].
\end{align}
We now invoke the EKF technique for predicting and tracking Bob's real-time location following the standard procedure. The state prediction and tracking can be summarized as follows:
 
1) \textit{State Prediction:} 
\begin{align}
  \label{stateprediction}
  \hat{\mathbf{x}}[n | n-1]=\boldsymbol{\Phi}[n|n-1]\hat{\mathbf{x}}[n-1].
\end{align}

2) \textit{Linearization:} 
\begin{align}
  \mathbf{H}[n]=\left.\frac{\partial h}{\partial \mathbf{x}}\right|_{\mathbf{x}=\hat{\mathbf{x}}[n | n-1]}.
\end{align}

3) \textit{MSE Matrix Prediction:} 
\begin{align}
 \quad \mathbf{P}[n | n-1]=\boldsymbol{\Phi}[n | n-1] \mathbf{P}[n-1] \boldsymbol{\Phi}^\text{T}[n | n-1]+\mathbf{Q}_{\omega}. \label{msepre}
\end{align}

4) \textit{Kalman Gain Calculation:}
\begin{align}
\mathbf{K}[n]= \mathbf{P}[n | n-1] \mathbf{H}^\text{T}[n] 
 \times\left(\mathbf{H}[n] \mathbf{P}[n | n-1] \mathbf{H}^\text{T}[n]+\sigma_d^2\right)^{-1}. \label{gain}
\end{align}

5) \textit{State Update:}
\begin{align}
  \label{update}
  \hat{\mathbf{x}}[n]=\hat{\mathbf{x}}[n | n-1]+\mathbf{K}[n](\hat{d}_b[n]-h(\hat{\mathbf{x}}[n | n-1])). 
\end{align}

6) \textit{MSE Matrix Update:} 
\begin{align}
  \label{mseupdate}
  \mathbf{P}[n]=(\mathbf{I}-\mathbf{K}[n] \mathbf{H}[n]) \mathbf{P}[n | n-1].
\end{align}
Note that $\hat{d}_b[n]$ in (\ref*{update}) is measured relying on the UAV's real-time location, which is derived in Section \ref*{real}.

\subsection{Real-Time Trajectory Design} \label{real}

 In this section, we solve the UAV's trajectory design problem for maximizing the real-time secrecy rate. Since the objective function (\ref*{optt1}) is non-convex with respect to $\mathbf{q}[n]$, the problem  (\ref*{optt}) is neither a convex nor a quasi-convex optimization problem, hence imposing challenges in finding the globally optimal solution. To overcome the non-convexity in (\ref*{optt1}), in the following, we harness the successive convex optimization method to obtain a sub-optional solution. We observe that the term $\Vert \mathbf{q}[n]-\mathbf{b}[n]\Vert^2$ is convex with respect to $\mathbf{q}[n]$ and the first part of (\ref*{optt1}) is convex with respect to $\Vert \mathbf{q}[n]-\mathbf{b}[n]\Vert^2$ as well. At this point recall that any convex function has a lower-bound given by the first-order Taylor series expanded at any point within its support. Let now $\mathbf{q}^r[n]$ denote the UAV's location in the $r$-th iteration at the $n$-th TS. Based on the predicted location $\hat{\mathbf{b}}[n|n-1]$ (included in $\hat{\mathbf{x}}[n|n-1]$), the lower-bound can be written as 

  \begin{align} 
    \label{first_order}
      \nonumber R_b[n]
      \nonumber &= B\log_2\left(1+\frac{p_0\rho_0}{\sigma^2\left(H^2+\Vert\mathbf{q}[n]-\mathbf{\hat{b}}[n\vert n-1]\Vert^2\right)}\right) \\ 
   \nonumber&\ge B\left(\log_2\left(1+\frac{p_0\rho_0}{\sigma^2\left(H^2+\Vert\mathbf{q}^r[n]-\mathbf{\hat{b}}[n\vert n-1]\Vert^2\right)}\right)\right.    \\
   \nonumber &+ \frac{1}{\ln 2}\left(\frac{1}{1+\frac{p_0\rho_0}{\sigma^2\left(H^2+\Vert\mathbf{q}^r[n]-\mathbf{\hat{b}}[n\vert n-1]\Vert^2\right)}}\right)\left(\frac{-p_0\rho_0}{\sigma^2}\right) \\
   \nonumber  & \left.\times \!\frac{\Vert\mathbf{q}[n]-\mathbf{\hat{b}}[n\vert n-1]\Vert^2-\Vert\mathbf{q}^r[n]-\mathbf{\hat{b}}[n\vert n-1]\Vert^2}{(H^2+\Vert\mathbf{q}^r[n]-\mathbf{\hat{b}}[n\vert n-1]\Vert^2)^2}\right)  \\
     &\triangleq \hat{R}_b[n]. 
  \end{align}
\par Next, by introducing the slack variables $\mathbf{s}[n]=\Vert\mathbf{q}[n]-\mathbf{w}[n]\Vert^2$, problem (\ref*{optt}) can be rewritten as
\begin{subequations}
  \label{cvx}
  \begin{align} \max_{\mathbf{q}[n],\mathbf{s}[n]} \quad  &\alpha \hat{R}_b[n] -(1-\alpha)B\log_2\left(1+\frac{p_0\rho_0}{\sigma_e^2\left(H^2+\mathbf{s}[n]\right)}\right) \label{obj}\\
  \text{s.t.} \quad & \mathbf{s}[n] \le \Vert\mathbf{q}[n]-\mathbf{w}[n]\Vert^2, \label{sn}\\
 &  (\ref*{velocity}),(\ref*{area}),(\ref*{area1}). 
\end{align}
\end{subequations}
\renewcommand{\algorithmicrequire}{\textbf{Initialization}} 
\renewcommand{\algorithmicensure}{\textbf{Output:}} 
\renewcommand{\thefootnote}{1}
\begin{algorithm}[t]
  \caption{ The Proposed Overall Algorithm} %
  \label{alg::conjugateGradient}
  \begin{algorithmic}[1]
    \State
      Initialize:set $\mathbf{x[1]},\mathbf{P}[1],\mathbf{q}[1]$, the corresponding index $n=1$, $r=1$, the tolerance $\epsilon$, the maximum iteration $r_{max}$.
    \Repeat  
      \State Set $n=n+1$. 
      \State  Compute  $\mathbf{\hat{x}}[n\vert n-1]$ with (\ref{stateprediction}) and $\mathbf{P}[n\vert n-1]$ with (\ref{msepre}).        
      \Repeat   
       \State With $\mathbf{\hat{b}}[n\vert n-1]$, $\mathbf{q}^r[n]$ to obtain $\mathbf{q}[n]$,  \text{obj} $(\mathbf{q}[n])^r$ \footnotemark[3]\hspace*{1cm} by solving problem (\ref*{cvx1}). 
       \State Update $r=r+1$.          
      \Until{$\vert$(obj$(\mathbf{q}[n])^{r}$-obj$(\mathbf{q}[n])^{r-1}$ $\vert\le \epsilon$ or $r>r_{max}$)}
      \State With $\mathbf{\hat{b}}[n\vert n-1]$ to obtain $\mathbf{K}[n]$ by (\ref{gain}).
      \State  With $\hat{d}_b[n]$ to obtain $\mathbf{\hat{x}}[n],\mathbf{P}[n]$ by (\ref{update}) and (\ref{mseupdate}).
    \Until{($n>N$)} 
  \end{algorithmic}
\end{algorithm} 
\footnotetext[3]{For notational convenience, we use obj$(\mathbf{q}[n])^r$ to denote the solution of problem (\ref*{cvx1}) in the $r$-th iteration.}

\noindent Although the objective function (\ref*{obj}) is now joint concave with respect to $\mathbf{q}[n]$ and $\mathbf{s}[n]$, problem (\ref*{cvx}) is still non-concave due to the constraint (\ref*{sn}). We find furthermore that $\Vert\mathbf{q}[n]-\mathbf{w}[n]\Vert^2$ is also lower-bounded by $\left\|\mathbf{q}^{r}[n]-\mathbf{w}[n]\right\|^{2}+2\left(\mathbf{q}^{r}[n]-\mathbf{w}[n]\right)^\text{T}\left(\mathbf{q}[n]-\mathbf{q}^{r}[n]\right)$. Hence, problem (\ref*{cvx}) can be  approximated as the following problem:
 \begin{subequations}
  \label{cvx1}
  \begin{align}  \max_{\mathbf{q}[n],\mathbf{s}[n]} \quad  &\alpha \hat{R}_b[n] -(1-\alpha)B\log_2\left(1+\frac{p_0\rho_0}{\sigma_e^2\left(H^2+\mathbf{s}[n]\right)}\right) \label{obj1}\\
  \text{s.t.} \quad  \nonumber &\mathbf{s}[n]\leq\left\|\mathbf{q}^{r}[n]-\mathbf{w}[n]\right\|^{2}\\& \qquad+2\left(\mathbf{q}^{r}[n]-\mathbf{w}[n]\right)^\text{T}\left(\mathbf{q} [n]-\mathbf{q}^{r}[n]\right), \label{sn1}\\
 & (\ref*{velocity}),(\ref*{area}),(\ref*{area1}). 
\end{align}
\end{subequations}
  Problem (\ref*{cvx1}) is concave, which can be readily solved using standard solvers, such as CVX \cite{grant2008cvx}. To sum up, the details of the proposed procedure are given in \textbf{Algorithm 1}. The optimal value increases with the iteration index, which is guaranteed to converge in a finite number of iterations \cite{convergent}.
     We further analyse the complexity of \textbf{Algorithm 1}. At each time slot, the EKF methods have to perform matrix inversion with a cubic complexity order of $\mathcal{O}(4^3)$. Then, upon denoting the iteration index of solving Problem (30) at time slot $n$ by $r_{n} (r_n\le r_{max})$ and assuming that the convex optimization Problem (30) is solved via the standard interior-point method having a complexity order of $\mathcal{O}[2^{3.5}\log(\frac{1}{\epsilon})]$ \cite{lu2022degrees}, the overall complexity of \textbf{Algorithm 1} is on the order of $\sum_{n=1}^N \limits \mathcal{O}\left[4^3+r_n(2^{3.5}\log(\frac{1}{\epsilon}))\right]$.

\section{ Numerical Results}

In this section, we present numerical results for characterising the performance of the proposed algorithm.  To guarantee the tracking performance while relying on a single UAV, we divide the entire duration into several tracking periods and  Bob will feed back his location to the UAV through the uplink channel at the beginning of each tracking  period. Here, we set each tracking period to $10$ TSs.  Note that the feedback mechanism indicates that the EKF tracking is employed within each tracking period respectively, rather than across the entire duration. Furthermore, the feedback mechanism can help us reliably distinguish Bob.  The size of region $\mathcal{D}$ is set as $L_x = L_y=1000$ m. We use $\Delta t=0.1$ s as the TS duration. The initial position of the UAV is the midpoint between Bob and Eve at the altitude of $H=50$ m and the maximum speed is $V_{max}=50$ m/s. The channel's power gain at a unit distance is set to $\rho_0=-60$ dB and the noise power in the receiver is set to $\sigma^2=\sigma_e^2= -100$ dBm.  We set the channel bandwidth as $1$ MHz.  The state transition noises are set with standard deviations $\sigma_x=\sigma_y=1$ m and $\sigma_{v_x}=\sigma_{v_y}=0.5$ m/s. Finally, for the observation, the standard deviations is set to $\sigma_{d}=2$ m.
 \begin{figure*}[t]
	 \centering  
	\subfigure[Received rates at Bob]{
	\includegraphics[width=0.46\linewidth]{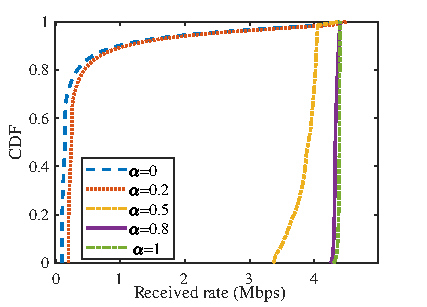}\label{DTRS}}
	\subfigure[Achievable secrecy rates]{
\includegraphics[width=0.46\linewidth]{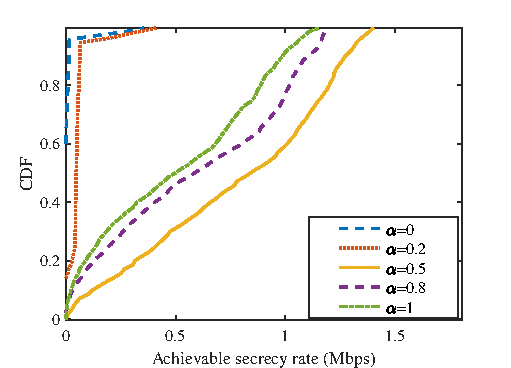}
		\label{serate}}
\caption{CDF of the throughput parameterized by $\alpha$.}
	\label{ratea}
 
\end{figure*}
 \begin{figure}[tb]
     \centering
     \includegraphics[width=0.45\textwidth]{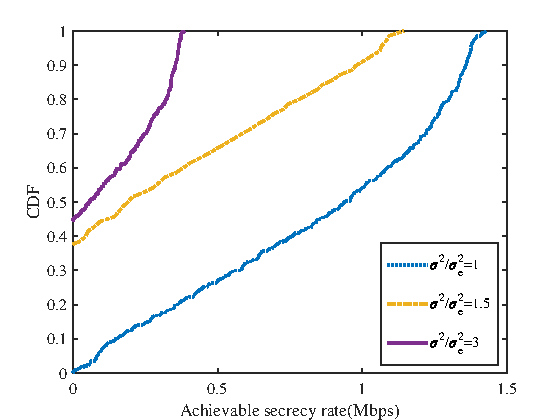}
     \caption{CDF of secrecy rate at different noise levels with $\alpha=0.5$.}
     \label{noise}
     \vspace{-0.5cm}
 \end{figure}

 \par
\begin{figure}[t]
  \centering
  \includegraphics[width=0.35\textwidth]{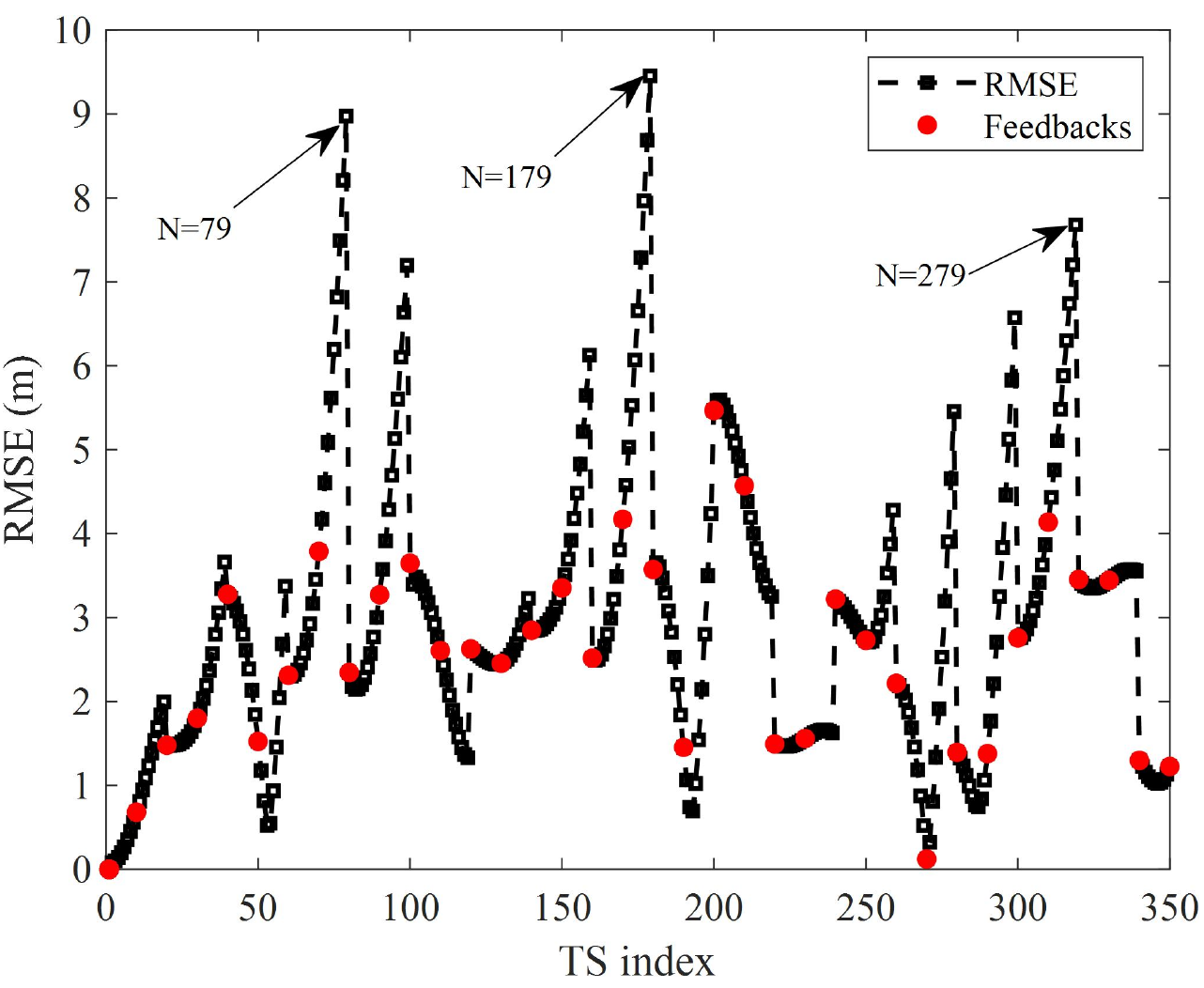}
   \captionsetup{singlelinecheck=off}
  \caption{ The RMSE of location estimation.}
  \label{errorpicture}
\end{figure}

\begin{figure}[t]
  \centering
\includegraphics[width=0.41\textwidth]{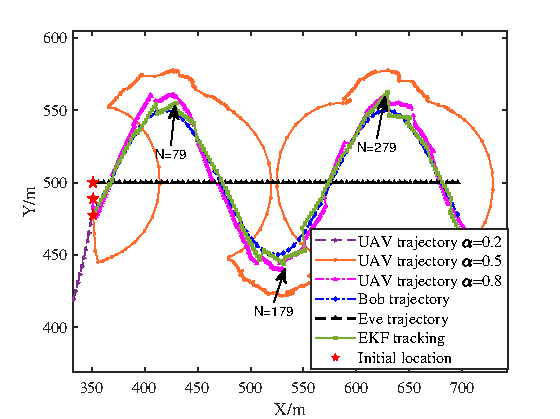}
  \captionsetup{singlelinecheck=off}
  \caption{The UAV's real-time trajectory parameterized by $\alpha$.}
  \label{trajectory}
\end{figure}

 In Fig. \ref*{ratea}, we first evaluate the cumulative distribution function (CDF) of both the legitimate received rate at Bob and the secrecy rate parameterized by $\alpha$.  It can be observed in Fig. \ref*{DTRS} that as the weighting factor $\alpha$ increases, the rate received at Bob also increases since the UAV trajectory will approach Bob's trajectory. By contrast, it is interesting to see in Fig. \ref*{serate} that although the scheme having $\alpha=0$ achieves the minimum leakage data rate, and the scheme with $\alpha=1$ achieves the maximum rates received at Bob, the proposed algorithm associated with $\alpha=0.5$ always achieves the best secure communication performance. To further investigate the effectiveness of our proposed algorithm, we present the CDF of the secrecy rate at different noise levels in Fig. $3$. As shown, we observe that as expected the secure communication performance is detrimentally affected by the noise ratio $\sigma^2/\sigma_e^2$, which is due to the fact that the eavesdropping capability of Eve is enhanced in the presence of lower noise levels. Nevertheless, our proposed algorithm supports secure communication most of the time, particularly when the ratio obeys $\sigma^2/\sigma_e^2\le 3$. Although the initial secrecy rate is zero, it gradually increases as \textbf{Algorithm 1} proceeds. When the secrecy rate is zero, we suspend the transmission of information.  However, it should be emphasized that when the ratio is too high,  the secrecy rate remains zero for the entire duration. 
We then study the tracking performance of the proposed EKF algorithm with Bob's initial location being $\mathbf{b}[1]=[350,470]^\text{T}$m and velocity being $\mathbf{v}[1]=[10,10]^\text{T}$m/s. In Fig. \ref*{errorpicture}, we evaluate the location estimation performance in terms of its root mean squared error (RMSE). Although the RSME increases in most cases after feedback due to the low positioning accuracy based on a single range measurement, it can be corrected with the aid of the location information from the following uplink feedback. Observe in Fig. \ref*{errorpicture}, the RMSE becomes high at specific TSs, e.g., the $79$-th TS, which is caused by the location prediction and estimation errors of Bob due to the movement direction change. 
 In Fig. \ref*{trajectory}, we show the UAV's real-time designed trajectory for different values of $\alpha$, where a higher value of  $\alpha$ indicates that the UAV focuses more on increasing Bob's communications rate rather than minimizing the leakage to Eve. As a result, the UAV trajectory almost coincides with Bob's trajectory. By contrast, we can see that the UAV will escape from Eve to the boundary of area $\mathcal{D}$ when $\alpha=0.2$, where the reduction of leakage rate is the main design goal. Furthermore, it can be observed that the UAV always endeavors to fly to the side far from Eve at each TS when $\alpha=0.5$, hence improving the security. On the one hand, the UAV will get close to Bob for improving the legitimate communications performance, which increases the risk of eavesdropping as well. On the other hand, the UAV flies away from Eve for reducing the leakage rate, while simultaneously reducing the legitimate communication rate for Bob. These trends motivate us to explore a more flexible secure communications mechanism.


\vspace{-0.2cm}
\section{Conclusion}
\par A UAV-aided dynamic ISAC system was designed, where a UAV having sensing and communication functionalities is supporting a ground user roaming through unknown locations in the presence of an eavesdropper. Specifically, we utilized the EKF algorithm for predicting and tracking the user's real-time location relying on the range measurements extracted from the ISAC echoes. Motivated by maximizing the real-time secrecy rate, we formulated a weighted real-time trajectory design problem. Furthermore, to solve the resultant non-convex optimization problem,  we proposed an efficient iterative algorithm relying on the popular successive convex optimization technique. 
Our simulation results verified that the proposed algorithm tracks the user efficiently and achieves flexible secure communications performance in UAV-aided applications having different communication requirements.



\vspace{-0.2cm}
\bibliographystyle{IEEEtran}
\bibliography{citation}
\end{document}